\documentclass[prl,showpacs,showkeys,twocolumn,spanish,nofootinbib]{revtex4-1}
\usepackage[spanish]{babel}
\usepackage[latin1]{inputenc}
\usepackage{bm}
\usepackage{amssymb,amsmath,amsthm}
\usepackage{mathrsfs}

\begin{document}

\title{{\Large\emph{Monólogo de una gravitona en crisis de identidad}}\\ \vspace{1mm}
O sobre las Teorías Alternativas de Gravedad}

\author{Prado Mart\'in Moruno}
\email{pradomm@ucm.es}
\affiliation{Departamento de F\'isica Te\'orica y UPARCOS,\\ Universidad Complutense de Madrid, E-28040 Madrid, Spain}

\begin{abstract}
La era de la Cosmología de alta precisión ha evidenciado nuestra ignorancia sobre la composición del Universo. En este contexto se ha renovado el interés por las Teorías Alternativas de Gravedad. A través de la experiencia de una gravitona medida por la colaboración LIGO, conoceremos el marco conceptual de estas teorías, que proporcionan un esquema teórico en el que contrastar las predicciones relativistas y un posible escenario en el que describir la expansión acelerada del Universo sin asumir la existencia de componentes oscuros. Durante el amanecer de la astronomía de ondas gravitacionales, la gravitación ya promete ser uno de los campos de investigación más fascinantes del siglo XXI.

\vspace{-3.5mm}
\begin{flushright}
\emph{Contribución invitada para la Revista Española de Física.}
\end{flushright}
\end{abstract}


\maketitle

\section{Mi nacimiento}
Todo empezó en la oscuridad del cosmos. Nací durante la última fracción de segundo del proceso de coalescencia de dos agujeros negros de masas unas 29 y 36 veces la masa del Sol, que dio como resultado un agujero negro todavía mayor\footnote{Por cierto, estos agujeros negros son objetos astronómicos fascinantes donde los haya, ya que ni siquiera la luz es capaz de abandonar el horizonte que los caracteriza, en lugar de tener una superficie como las estrellas y planetas.} [1]. No vine sola. En una fracción de segundo una energía correspondiente a 3 veces la masa del Sol se convirtió en hordas de compañeras que iniciamos un viaje por la infinidad del Universo a la velocidad de la luz [2], estirando y comprimiendo el espacio a nuestro paso. Había viajado durante más de mil millones de años cuando pasé por un planeta bastante peculiar. Allí, en la Tierra, un equipo formado por 1.000 personas empezó a analizar datos y se alegraron mucho al concluir que habían notado nuestra presencia mediante dos detectores [1], lo que les permitía estar seguros de su hallazgo. Inicialmente me sorprendió el júbilo que demostraron los habitantes de ese planeta cuando les comunicaron la noticia. Los expertos del lugar incluso afirmaron que se había abierto una nueva ventana para observar el Universo, ya que antes sólo podían estudiarlo por medio de fotones que, al contrario de nosotras las gravitonas, pueden ser absorbidos o reflejados por la materia. 
Más tarde comprendí que los habitantes de ese plantea llevaban ya mucho tiempo hablando de nosotras e intentando encontrarnos. A principios del siglo XX un pionero en la comprensión del Universo, Albert Einstein, supo entender que el espacio y el tiempo forman una estructura espacio-temporal que se curva cuando contiene energía, afectando así al movimiento de la materia por esa misma estructura. La teoría desarrollada por Einstein permitió a los habitantes de la Tierra, entre otras cosas, describir el movimiento de los planteas con gran exactitud, entender el cambio de la trayectoria de la luz al pasar cerca de objetos muy densos e implicó el nacimiento de la Cosmología como ciencia. Sin embargo, según escuché, esa era la primera detección directa de ondas gravitacionales, confirmándose así la validez de la Teoría de la Relatividad General de Einstein en un fenómeno proveniente de un acontecimiento en el que la gravedad era muy fuerte, es decir, el espacio-tiempo estaba muy curvado.

\section{Asumamos mi existencia}
Yo no me siento tan especial. A veces me pregunto incluso si de verdad existo. A fin de cuentas los argumentos cartesianos no parecen aplicables cuando a una no le reconocen la capacidad de pensamiento. Las ondas gravitacionales son ondulaciones del espacio-tiempo, fácilmente identificables como pequeñas perturbaciones en un espacio-tiempo plano, lorentzianamente hablando. Sin embargo, no está tan claro como definir las ondas gravitacionales en geometrías curvas arbitrarias en las que la separación entre fondo y onda no es en general unívoca. Esto se podría hacer en situaciones donde haya dos escalas diferenciadas, una que caracterizara a la variación del espacio de fondo y la otra a la onda gravitacional [3]. Es más, cuando se habla de mí, la partícula asociada a la onda gravitacional, se supone implícita o explícitamente una cuantización de la perturbación vista como un campo definido en la geometría de fondo. Por lo tanto, a falta de una teoría de gravedad cuántica completa, mi existencia misma es debatible. De todos modos, me voy a tomar la licencia literaria de suponer que existo.

\section{Mi naturaleza}
Pero eso no es lo que quería contaros. Al pasar por la Tierra escuché algunas cosas sobre la naturaleza de los fenómenos gravitatorios que, de ser ciertas, afectarían mis propias cualidades. Ahora me pregunto cómo soy realmente. Según la teoría de la Relatividad General la gravedad es de naturaleza puramente geométrica, lo que ha sido confirmado por numerosos experimentos [4]. Citando a John A.~Wheeler ``El espacio-tiempo le dice a la materia cómo moverse; la materia le dice al espacio-tiempo cómo curvarse" [5]. La geometría del espacio-tiempo, descrita por su tensor métrico, es la solución proporcionada por las ecuaciones de la teoría cuando conocemos la materia presente. Así, el espacio-tiempo no es el lugar donde pasa la física sino un ente físico en sí mismo. Todo, incluso la luz, siente la gravedad, ya que ésta no es más que una propiedad de la geometría que alberga la realidad. 

\subsection{Sobre principios y teorías}
El principio de Mach, que se relaciona con la idea de que las propiedades del espacio tienen su origen en la materia que contiene, es acorde con este tipo de razonamientos relativistas. Sin embargo, en 1961 Carl H.~Brans y Robert H.~Dicke [6] reflexionaron sobre la posibilidad de que la Relatividad General no recogiera completamente todas sus implicaciones. Estos autores argumentaron que para que la aceleración de una partícula en un campo gravitatorio estuviera realmente determinada por la distribución de la materia en el Universo, la constante gravitatoria $G$ observada localmente no debería ser constante, sino que se tendría que ver afectada por la distribución de materia en cuestión. Para realizar esta idea formularon una teoría gravitatoria que incluía un campo gravitacional escalar además de la métrica de la geometría (que es un campo tensorial). 

	La teoría de Brans--Dicke es, sin lugar a dudas, el ejemplo más conocido de teoría de gravedad escalar-tensor de las muchas que se han considerado más tarde. Por otra parte, otros científicos también reflexionaron sobre que la teoría de la Relatividad General debería modificarse en presencia de energías muy grandes, para poder combinar la gravedad con el resto de interacciones conocidas. Siguiendo este espíritu investigaron teorías que extendían la teoría relativista construidas con los mismos ``ladrillos geométricos" de la teoría de Einstein, aunque alterando la relación entre materia y geometría [7]. Esos ladrillos son la regla de medida que nos proporciona la distancia entre dos puntos espacio-temporales (la métrica de la geometría) y la acción que nos permite comparar cantidades definidas en puntos distintos del espacio-tiempo (la conexión)\footnote{La regla y el compás representan la métrica, mientras que la escuadra y el cartabón representan las operaciones asociadas a la conexión.} . Aunque en Relatividad General es equivalente considerar la métrica y la conexión como dependientes o no, en otras teorías esa independencia generaría una realidad física completamente distinta. En general, todas las Teorías Alternativas de Gravedad reformulan el dialogo entre la materia y la geometría, pudiendo incluir a otros actores en la discusión (como, por ejemplo, el campo escalar de Brans y Dicke).

	Al igual que la Relatividad General las Teorías Alternativas de Gravedad, o la gran mayoría de las consideradas como potencialmente viables, son compatibles con el principio de equivalencia débil, según el cual el comportamiento de partículas de prueba (en la que los efectos de la gravedad que generan ellas mismas son despreciables) en un campo gravitatorio es independiente de sus propiedades. Sin embargo, hasta donde se sabe, sólo la Relatividad General cumple el principio de equivalencia fuerte, que postula la independencia local de toda la física de la posible presencia de un campo gravitacional [8]. 

\subsection{El rompecabezas cosmológico}
El mayor conocimiento sobre el cosmos ha llevado a los entendidos a poner otra vez en cuestión mis propiedades, esta vez no sólo por argumentos formales sino también fenomenológicos. En la actualidad la comunidad científica tiene a su disposición datos provenientes de las observaciones de una calidad sin precedentes. Sin embargo, la interpretación de esos datos en el marco de la Relatividad General les ha mostrado su gran ignorancia sobre la composición del Universo, señalando que desconocen la naturaleza del 95\% del contenido energético cósmico [9]. Este contenido misterioso debería ser oscuro (es decir, debería interaccionar a lo sumo débilmente con el campo electromagnético), ya que no lo ven de forma directa. Mientras que aproximadamente el 27\% del contenido cósmico puede formar acumulaciones, por lo que se conoce como materia oscura, el 68\% de ese contenido, llamado energía oscura, está distribuido uniformemente por el espacio y debe tener propiedades gravitatorias repulsivas. Se supone que esa gran cantidad de energía oscura es la responsable de que la expansión actual de nuestro Universo no se frene, si no que se acelere. El modelo cosmológico estándar actual asume que la energía oscura tiene una densidad de energía constante durante la evolución del Universo, es decir, se puede describir mediante un término de constante cosmológica, como el introducido y desechado por Albert Einstein hace un siglo para obtener un modelo de universo estático. Sin embargo, este término cosmológico, interpretable como la energía del vacío, lleva de cabeza a los científicos que no paran de plantear problemas con su nombre, relativos probablemente a su falta de pericia calculando su valor [10,11].

	Algunos investigadores han pensado que, como por el momento la física de partículas no les ha proporcionado información satisfactoria para comprender este rompecabezas cosmológico y la Relatividad General no ha sido comprobada en todas las escalas de forma apropiada, la energía oscura podría ser simplemente el resultado de ignorar posibles modificaciones de las predicciones relativistas que serían relevantes al considerar distancias muy grandes. A fin de cuentas estas teorías ya se investigaban para describir al Universo durante la otra época de expansión acelerada que se cree experimentó cuando era muy joven y energético, el periodo inflacionario; actualmente incluso se considera que el modelo inflacionario más prometedor se basa en una de estas teorías, el modelo de Starobinsky [9]. Así, podría ser que la Relatividad General fuera válida solamente a escalas intermedias y que la energía oscura y el inflatón fueran simplemente una señal de dinámica gravitatoria asociada a otra teoría que la englobase\footnote{En este caso, volveríamos a encontrarnos con el ``viejo problema de la constante cosmológica" [10] que consiste en explicar por qué el valor de este término se anula, o se apantalla mediante algún mecanismo de estas teorías (véase, por ejemplo [12,13]), dejando de tener efecto.}. 

\subsection{Cómo soy}
Esta fauna de teorías gravitatorias me han hecho plantearme mi propia naturaleza, reconsiderando qué sé y qué no sé de mí misma. Si la Relatividad General fuera realmente válida a todas las escalas, las ondas gravitacionales tendrían dos polarizaciones tensoriales transversas a la propagación de la onda que viajarían a la velocidad de la luz. Por lo tanto, mi comportamiento sería el de una partícula sin masa, como sucede en otras interacciones, y con spin 2 (recordad, por ejemplo, que los fotones son partículas sin masa con spin 1). 

	Sin embargo, si la teoría gravitatoria fuera otra alternativa, yo podría ser muy distinta. La radiación gravitatoria podría constar de hasta 6 modos de polarización, al poderse predecir 2 modos vectoriales y 2 escalares además de los 2 tensoriales. Por lo tanto yo podría estar formada por otras partes que viajaran siempre conmigo, ¡o me persiguiesen!. Es más, en principio es posible que algunas teorías que se consideran viables, al estar de acuerdo con las observaciones a varias escalas, predigan que me propago a una velocidad distinta que la de la luz, como pasa en teorías escalar-tensor complicadas (no en la de Brans--Dicke). Puestos a imaginar, podría hasta tener masa según las Teorías de Gravedad Masiva. Los estudiosos han concluido que esas teorías podrían formularse de forma apropiada, sin predecir el sexto modo escalar que introduce inestabilidades [14]. Sin embargo, en ese escenario, los modos vectoriales no se producirían por eventos como la coalescencia de agujeros negros (al no acoplarse con la materia), el único modo escalar podría apantallarse y sólo se me distinguiría de mi hipótesis relativista si viajara distancias muy grandes (mayores que mi longitud de onda Compton) [14].

	Por su puesto, no todo vale. Cualquier teoría que quiera disputarle a la Relatividad General su estatus de teoría gravitatoria estándar debe describir de forma satisfactoria todas las predicciones de ésta que ya han sido comprobadas, que son muchas [4]. Al igual que la teoría de Einstein, esa posible candidata deberá recuperar las predicciones de Newton en la Tierra y describir el movimiento de los planetas en el Sistema Solar, incluida la predicción estrella de la teoría einsteniana sobre el avance del perihelio de Mercurio (comprobaciones de este tipo se han refinado mucho gracias a los púlsares binarios). Esa teoría deberá además predecir la física de las galaxias y grupos de galaxias, probablemente asumiendo también una materia oscura elusiva que podría dejar de esconderse a las observaciones de los científicos en un futuro cercano. También tendrá que recuperar las predicciones relativistas sobre la evolución del Universo durante la mayor parte de su historia, salvo tal vez sus primeras fracciones de segundo de vida y aproximadamente durante los últimos 5 mil millones de años, en los que lo ideal sería que describiera la expansión acelerada asumiendo sólo la existencia de materia y radiación. Todas estas predicciones han implicado que se limitara en gran medida el rango de valores que pueden tomar los parámetros de las Teorías Alternativas de Gravedad  bajo estudio, descartándose algunas candidatas. Además, hace menos de un año los científicos han sido capaces de medir el paso de compañeras gravitonas que se formaron en un fenómeno que también emitió señales luminosas [2]. Estas mediciones han evidenciado que nuestra velocidad de propagación en la actualidad y recientemente (cosmológicamente hablando) debe ser la misma que la de la luz, lo que descarta que muchas Teorías Alternativas de Gravedad sean buenas candidatas para describir la época de expansión acelerada sin energía oscura [15].

\subsection{Desvelando el misterio}
En la actualidad, hay muchos investigadores intentando formular y descartar Teorías Alternativas de Gravedad y modelos de energía oscura dinámica. Por un lado, los observatorios y misiones espaciales les proporcionarán más información sobre la confirmación o reconsideración de la teoría de la Relatividad General. Por otro lado, también están desarrollando distintos marcos teóricos para comprender pruebas gravitatorias adicionales, como formas en las que las gravitonas podremos seguir ayudando a conocer mejor el Universo.

	¿De verdad debemos confiar la dinámica cósmica a un nuevo éter, o existe una teoría capaz de describir los fenómenos gravitatorios en todas las escalas observables sin necesidad de energía oscura? Creo que la convicción de haber encontrado respuesta a esta pregunta sería la única forma de conocerme realmente. Tal vez los investigadores ya tengan a su disposición la teoría gravitatoria última (y ésta sea la Relatividad General) o, en mi opinión, es muy probable que debamos inspirarnos en Niels Bohr y concluir que ninguna teoría de las candidatas actuales es lo suficientemente disparatada para ser cierta.

\bigskip
\noindent {\bf Aviso importante}: No se ha dañado a ninguna gravitona durante la redacción de este artículo. Lo he intentado, pero no he sido capaz.

\bigskip
\noindent \emph{\bf Agradecimientos:}
Agradezco los fondos recibidos a través del premio L'Oréal-UNESCO For Women in Science (XII edición española).

\section{Bibliografía}
\noindent [1] B. P. Abbott et al. [LIGO Scientific and Virgo Collaborations], ``Observation of Gravitational Waves from a Binary Black Hole Merger", Phys. Rev. Lett. 116, 061102 (2016). 

\noindent [2] B. P. Abbott et al. [LIGO Scientific and Virgo Collaborations], ``GW170817: Observation of Gravitational Waves from a Binary Neutron Star Inspiral", Phys. Rev. Lett. 119 (2017) no.16, 161101.

\noindent [3] C. Caprini and D. G. Figueroa, ``Cosmological Backgrounds of Gravitational Waves", arXiv:1801.04268 [astro-ph.CO].

\noindent [4] C. M. Will, ``The Confrontation between General Relativity and Experiment", Living Rev. Relativity 17 (2014) 4.

\noindent [5] J. A. Wheeler, ``Geons, Black Holes, and Quantum Foam", Norton \& Co Inc. 2000.

\noindent [6] C. H. Brans and R. H. Dicke, ``Mach?s Principle and Relativistic Theory of Gravitation", Phys. Rev. 124 (1961) 925.

\noindent [7] S. Capozziello and M. De Laurentis, ``Extended Theories of Gravity", Phys. Rept. 509 (2011) 167.

\noindent [8] E. Di Casola, S. Liberati, and S. Sonego, ``Nonequivalence of equivalence principles", Am. J. Phys. 83 (2015) 39.

\noindent [9] P. A. R. Ade et al. [Planck Collaboration], ``Planck 2015 results. XIII. Cosmological parameters", Astron. Astrophys. 594 (2016) A13.

\noindent [10] S. Weinberg, ``The Cosmological Constant Problem", Rev. Mod. Phys. 61 (1989) 1.

\noindent [11] V. Sahni, ``The Cosmological constant problema and quintessence", Class. Quant. Grav. 19 (2002) 3435.

\noindent [12] N. Kaloper and A. Padilla, ``Sequestering the Sandard Model Vacuum Energy", Phys. Rev. Lett. 112 (2014) 091304.

\noindent [13] P. Martín-Moruno, N. J. Nunes, and F. S. N. Lobo, ``Horndeski theories self-tuning to a de Sitter vacuum", Phys. Rev. D91 (2015) 084029.

\noindent [14] C. de Rham, ``Massive Gravity", Living Rev. Rel 17 (2014) 7.

\noindent [15] J. M. Ezquiaga and M. Zumalacárregui, ``Dark Energy After GW170817: Dead Ends and the Road Ahead", Phys. Rev. Lett. 119 (2017) 251304.

\end{document}